\documentclass[fleqn,twocolumn, superscriptaddress]{revtex4}
\usepackage{epsfig}
\usepackage{graphicx}
\usepackage{color}
\usepackage{amsfonts,amssymb}

\newcommand{\ct}{\cite}
\newcommand{\bi}{\bibitem}
\newcommand{\be}{\begin{equation}}
\newcommand{\ee}{\end{equation}}
\newcommand{\ba}{\begin{eqnarray}}
\newcommand{\ea}{\end{eqnarray}}

\newcommand{\non}{\nonumber}

\begin{document}
\title{Efficiency of quantum controlled non-Markovian thermalization}

\author{V.~Mukherjee}
\affiliation{NEST, Scuola Normale Superiore \& Istituto Nanoscienze-CNR, I-56126 Pisa, Italy}
\affiliation{
Institute for Complex Quantum Systems \& IQST, Universit\"at Ulm, D-89069 Ulm, Germany}
\author{V. Giovannetti}
\affiliation{NEST, Scuola Normale Superiore \& Istituto Nanoscienze-CNR, I-56126 Pisa, Italy}
\author{R. Fazio}
\affiliation{NEST, Scuola Normale Superiore \& Istituto Nanoscienze-CNR, I-56126 Pisa, Italy}
\affiliation{Centre for Quantum Technologies, National University of Singapore, 117543, Singapore}
\author{S.~F. Huelga}
\affiliation{
Institut f\"ur Theoretische Physik \& IQST, Universit\"at Ulm, D-89069 Ulm, Germany}
\author{T.~Calarco}
\affiliation{
Institute for Complex Quantum Systems \& IQST, Universit\"at Ulm, D-89069 Ulm, Germany}\author{S.~Montangero}
\affiliation{
Institute for Complex Quantum Systems \& IQST, Universit\"at Ulm, D-89069 Ulm, Germany}

\begin{abstract}
We study optimal control strategies to optimize the relaxation rate towards the fixed point of a quantum system in the presence of a
non-Markovian dissipative bath. Contrary to naive expectations that suggest that memory effects might be exploited to improve
optimal control effectiveness, non-Markovian effects influence the optimal strategy in a non trivial way: we present a necessary condition
to be satisfied so that the effectiveness of optimal control is enhanced by non-Markovianity subject to suitable unitary controls. 
For illustration, we specialize our findings for the case of the dynamics of single qubit amplitude damping channels.
The optimal control strategy presented
here can be used to implement optimal cooling processes in quantum technologies and may have implications in quantum thermodynamics when assessing the efficiency of thermal micro-machines.
\end{abstract}

\maketitle

\section{Introduction}
\label{intro}
Controlling quantum systems by using time-dependent  fields~\cite{krotov} is of primary importance in different branches
of science, ranging from  chemical reactions~\ct{Zewail80, brif10}, NMR~\cite{stefanatos04}, molecular physics~\ct{tannor99}
to the emergent quantum technologies~\ct{rabitz93, campo12, wamsley}.
Investigations on optimal control of open quantum systems mostly focus on memoryless environments ~\ct{sugny07, sauer13, haas14} and specifically on those situations where the reduced dynamics can be described by a Markovian master equation of the Lindblad form \cite{LKGS}. In this context, optimal control applications to open quantum systems have been explored in different settings~\cite{tannor99,lloydviola,rebentrost,hwang,roloff,
carlini06,carlini08} and recently the ultimate limits to optimal control dictated by quantum mechanics in closed and open systems~\cite{caneva09, campo13,taddei13,heydari13}
and the complexity of dealing with many-body systems~\ct{doria11, caneva14, lloyd14} have been determined.
Time-optimal quantum control has been extensively discussed for one qubit systems in a dissipative environment~\ct{sugny07,lapert10}
and the optimal relaxation times determined in~\cite{mukherjee13}. These studies might have both fundamental and practical applications, for example
in assessing the ultimate efficiency of quantum thermal machines ~\cite{kosloff}, or to implement fast cooling schemes which have already proven to be advantageous~\ct{machnes10, hoffmann}.\\
However, introducing a Markovian approximation requires some constraints on system and environment, which may not be valid in general~\ct{breuer03, huelga11}.
Consequently, incorporating non-Markovian (NM) effects of the environment, in a sense that will be defined more precisely below, might be a necessity in a many experimental situations.
Recently, the possible influence of memory effects on the orthogonality catastrophe \cite{sindona13}, on quantum speed of evolution \cite{deffner13}  and on quantum control \ct{koch14, maniscalco15} have been analyzed.
Here, we present a study of the optimal control strategies to manipulate quantum systems in the presence of NM dissipative baths and compare
the performance of optimal control with the case of operating subject to a Markovian (M) environment.

Intuitively, the absence of memory effects in the dynamics of open quantum systems is linked to the possibility of identifying well separated time scales in the evolution of system and environment. Recently, a number of proposals 
have been put forward to quantitatively characterize this effect in terms of explicit non-Markovianity measures \cite{apollaro11, lorenzo13, plenio14,addis14}. In this light, one can define an evolution to be Markovian if described by a quantum dynamical 
semigroup (time-homogeneous Lindbladian evolution) \cite{eisert}, which would be the traditional Markovianity considered in most previous work on open system control. However, other definitions encompass this as a special case while allowing 
for more 
general, non-homogeneous generators, albeit still ensuring the  divisibility of the associated dynamical map and the unidirectionality of the system-environment information flow, and therefore the absence of memory
effects in the dynamics of the system \ct{rivas11, santos14, wudarski15, santos15}. Relevant for our analysis is the
definition
of Markovian evolution in terms of the divisibility of the associated dynamical map \cite{huelga10}. When the dynamics is parametrized using a time-local master equation, the requirement of trace and hermiticity preservation, yields a generator of the
form
\begin{eqnarray}\label{FIRST}
\dot{\rho_s}(t) 
&=& -{i}\left[H_s(t),\rho(t)\right]+ \bar{\cal{L}}(t)(\rho(t)) \non\\ &=& -{i}\left[H_s(t),\rho(t)\right] + \sum_k \gamma_k(t) (A_k(t)\rho(t)A_k^{\dagger}(t)\\ \nonumber &-&\frac{1}{2}\{A_k^{\dagger}(t)A_k(t) \rho(t)\}),
\end{eqnarray}
where $\bar{\cal{L}}(t)$ is a time dependent Lindblad superoperator, $\gamma_k(t)$ are generalized (i.e. not necessarily positive) decay rates, the $A_k(t)$'s form an orthonormal basis for the operators for the system, see e.g.
Ref.~\cite{hall14} (hereafter $\hbar$ has
been set equal to one for convenience) and $H_s(t)$ is the effective Hamiltonian acting on the system. 
Equation~(\ref{FIRST})  generalizes the familiar Lindbladian structure 
to include NM effects while maintaining a time-local structure.
However, apart from same special cases,  it not known which are the conditions which 
$H_s(t)$, $A_k(t)$, and  $\gamma_k(t)$ have to satisfy in order to guarantee Complete Positivity (CP)~\cite{hall14,Kossakowski10,hall08,won,sab}, i.e. the fundamental prerequisite which under fairly general assumptions is needed to describe a
proper quantum evolution~\cite{breuer03,huelga11}.
In what follows we will focus on a simplified scenario where the  $\gamma_k(t)$'s either are null or coincide with an assigned function $\gamma(t)$, and where 
the $A_k(t)$'s are explicitly time-independent. Accordingly, in the absence of any control Hamiltonian applied during the course of the evolution, we assume a dynamical evolution described by the equation
\ba
\dot{\rho}(t) = \gamma(t) {\cal L} (\rho(t))\;,
\label{mastergen}
\ea
where ${\cal L}$ is a (time-independent) Lindblad generator characterized by having a unique fixed point $\rho_{fp}$ (i.e. ${\cal L}(\rho) = 0$
iff $\rho= \rho_{fp}$).
For this model, in the absence of any Hamiltonian term
(i.e. $H_s(t) =0$) CP over a time interval $[0,T]$  is guaranteed when~\ct{Kossakowski10} 
\begin{eqnarray}
\int^{t}_0 \gamma(t') dt' \geq 0\;, \qquad \forall t\in[0,T]\;,  \label{JNEWCON}
\end{eqnarray}
while divisibility (i.e, Markovianity) is tantamount to the positivity of the single decay rate at all times \cite{hall14}: if there exists a time interval where $\gamma(t)$ becomes negative, the ensuing dynamics is no longer 
divisible and the evolution is NM. 
 In this context  we will assume a control Hamiltonian $H_s(t)$ to represent time-localized infinitely strong pulses, which 
induce instantaneous unitary transformations at specific control times. This corresponds to writing 
$H_s(t) = \sum_{j} \delta(t-t_j)  \Theta_s^{(j)}$, 
where $\Theta_s^{(j)}$ are time independent operators which act impulsively on the system at $t=t_j$ ($\delta(t)$ being the Dirac delta-function), at which instants one can neglect the contribution from the non-unitary part, and 
represent the master equation by $\dot{\rho}(t)
\approx -{i}\left[H_s(t),\rho(t)\right]$. 
Therefore the resulting dynamics is described by a sequence of free evolutions induced by the noise   over the intervals   $t\in [t_j , t_{j+1}]$ 
interweaved with unitary rotations $U_{j} = \exp[-i \Theta_s^{(j)}]$, i.e. 
\begin{eqnarray} \label{JNEWEQ} 
\rho(t) &=&{\cal U}_{fin} \circ  {\cal D}_{j} \circ {\cal U}_{j} \circ {\cal D}_{j-1} \circ \cdots \nonumber \\
&&\qquad \qquad \cdots  \circ {\cal U}_{1}\circ {\cal D}_{0} \circ {\cal U}_{in}(\rho(0))\;,
\label{fullevol}
\end{eqnarray} 
where ${\cal D}_{j} = \exp\left[ \int^{t_{j+1}}_{t_j}  \bar{\cal L}(t)\right]$, ${\cal U}(\cdots) = U (\cdots) U^{\dag}$ and ``$\circ$" is the composition of super-operators.
When
only two control pulses are applied (the first ${\cal U}_{in}$ at the very beginning and the second ${\cal U}_{fin}$ at the very end of the temporal evolution), the non-unitary evolution is described by Eq. (\ref{mastergen})  and 
CP of the trajectory (\ref{fullevol}) is automatically guaranteed by Eq.~(\ref{JNEWCON}),
the scenario corresponding to the realistic case where one acts on the system with very strong control pulses at the state preparation stage and immediately before
the measuring stage.  When more ${\cal U}_{j}$'s are present, the situation however  becomes  more complex. There is no clear physical prescription which one can follow to impose the associated dynamics on the system at least when the 
dissipative evolution is assumed to be NM.  Consider for instance the case ${\cal U}_{fin} \circ  {\cal D}_1  \circ {\cal U}_{1}\circ {\cal D}_{0} \circ {\cal U}_{in}(\rho(0))$ where  ${\cal U}_1$  is a non trivial unitary. 
Even admitting that the latter is enforced by applying at time $t_1$ a strong instantaneous control pulse, there is absolutely no clear evidence that the open dynamics for $t\geq t_1$ should be
still described by  the same generalized Lindbladian $\gamma(t) {\cal L}$, the system environment being highly sensitive to whatever the system itself has experienced in its previous history. 

We note that in a more realistic scenario, any control pulse will have a non-zero width $\delta t$ in time. Clearly, a sufficiently large $\delta t$ can invalidate the assumption of applying control pulses only 
at the very beginning and the very end, thereby modifying the dynamics significantly as described above.
However, one can expect the assumption of instantaneous pulses 
to be valid as long as $\delta t$ is negligible compared to the time scale associated with the dynamics in absence of any control.

Keeping in mind the above limitations, in this work we attempt for the first time a systematic study of 
optimal control protocols which would allow one to  speed up the driving of a generic (but known) initial state $\rho(0)$ toward 
the fixed point~$\rho_{fp}$  of the bare dissipative evolution for the model of Eq.~(\ref{mastergen}) which explicitly includes NM effects. We arrive at the quantum speed limit times when application of only
two control pulses ${\cal U}_{in}$ and ${\cal U}_{fin}$,
at initial and final times respectively, is enough to follow the optimal trajectory.
On the other hand, we present lower bounds to the same when optimal control strategy demands unitary pulses at intermediate times as well.
We show that the efficiency of optimal control protocols
 is not determined by the M/NM divide alone but it depends drastically on the behaviour of the NM channel: if the system displays NM behavior 
{\it before} reaching the fixed point for the first time, NM effects might be exploited to
obtain an increased optimal control efficiency as compared to the M scenario. On the contrary,
NM effects are detrimental to the optimal control effectiveness if information back-flow occurs only {\it after} the system reaches the fixed point (see Fig.~\ref{schematic}).
These results are valid irrespective of the detailed description of the system, i.e. its dimension, Hamiltonian, control field, or the explicit form of the dissipative bath.

\section{The model\label{Rgen}}
\begin{figure}[t]
\begin{center}
\includegraphics[width= 8 cm, angle = 0]{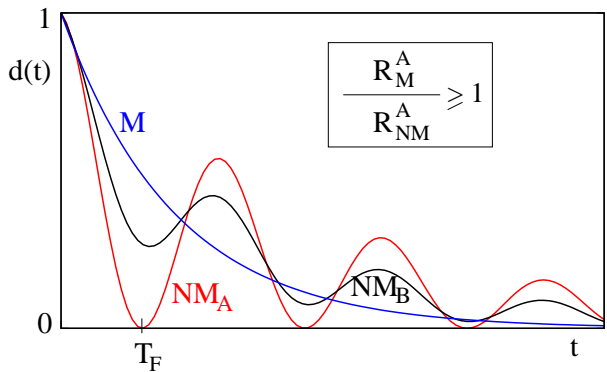}
\end{center}
\caption {(Color online) Schematic diagram of NM dynamics in Class A (red line) and Class B (black line). The instantaneous trace distance $d(t) = ||\rho(t) - \rho_{fp}||$
 starts increasing only after the system reaches the fixed point when $|\gamma_t| \to \infty$ in case of Class A, while it shows oscillatory behavior even before it reaches the fixed point in case of Class B.
 In comparison dynamics for a M channel is shown by the blue line where $d(t)$ decreases monotonically and assymptotically to $d(t \to \infty) = 0$. The speedup obtained by M
dynamics is always bigger than that obtained for the NM one, i.e., $R^A_{M}/R^A_{NM} \geq 1$ in case the NM evolution is of Class A, while the Markovian limit can be surpassed by NM of class B.}
\label{schematic}
\end{figure}

The divisibility measure for the model Eq.~(\ref{mastergen})  is equivalent to the characterization of memory effects by means of the time evolution of the trace distance \cite{breuer09}. This provides an 
intuitive characterization of the presence of memory effects in terms of a temporary increase in the distinguishability of quantum states as a result of an information back-flow from the system and into the environment that is
absent when the evolution is divisible \cite{breuer2}.
As a result, a divisible evolution for which the single decay rate $\gamma(t) \ge 0$ at all times will exhibit a monotonic decrease of the trace distance of any input state towards a (assumed to be unique) fixed point $\rho_{fp}$ of the
Lindblad generator ${\cal L}$ \cite{breuer3}. On the contrary, as illustrated in Fig.~\ref{schematic}, the behaviour of the trace distance can be non-monotonic when the dynamics is NM. In this case, there exist time interval(s) where $\gamma(t)$ becomes negative.
Denoting by $d(t) = ||\rho(t) -\rho_{fp}||$ the trace distance between $\rho(t)$ and the fixed point $\rho_{fp}$, it straightforwardly follows that $\dot{d}(t) \leq 0$ $\forall t$ in the M limit.
Looking at this quantity one can classify NM dynamics into two distinct classes (see Fig.~\ref{schematic}):
the first one (Class A) is defined by those dynamics where the system reaches the fixed point at time $T_{F}$ before $\gamma(t)$
changes sign i.e., $\gamma(t) \geq 0$ and $\dot{d}(t) \leq 0$ for $0 \leq t < T_{F}$. In this case, the NM dynamics
reaches the fixed point $\rho_{fp}$ and then start to oscillate. On the other hand, Class B dynamics is characterized by
$\gamma(t)$ that changes sign (and correspondingly $\dot{d}(t) > 0$) at some time $t < T_{F}$, that is the solutions of the equation $\gamma(t_s) = 0$ are such that $t_s < T_F$ for at least one $s$. In contrast, in the M dynamics
$d(t)$ always decreases monotonically and asymptotically to $d(t \to \infty) = 0$.

NM channels of class A/B arise from different physical implementations. As an illustration, the damped Jaynes Cummings model exemplifies a Class A dynamics. Here a qubit is coupled to a single cavity mode which in turn is coupled to a reservoir
consisting of harmonic oscillators in the vacuum state (see Eq. (\ref{JC}))\ct{breuer99, garraway97, breuer03, madsen11}. On the other hand,
dynamics similar to Class B can arise for example in a two level system in contact with an environment made of
another two level system, as realized recently in an experimental demonstration of NM dynamics \ct{souza13}.

As we will see hereafter, the difference between Class A and B appears to drastically affect the performance of any possible optimal control strategy to improve the speed
of relaxation of the system towards the fixed point.
\newline 

We assume full knowledge of the initial state and we  allow for an error tolerance of $0 < \epsilon \ll 1$, considering  that the target is reached
whenever the condition $|d(t)| \leq \epsilon$ is satisfied.
To obtain a lower bound on the minimum time $T_{QSL}$ needed to fulfill such constraint we restrict our analysis to the ideal limit of infinite control
which allows us to carry out any unitary transformation instantaneously  along the lines  of (and with all the limitations associated with) the formalism detailed in Eq.~(\ref{JNEWEQ}).
In the limit of infinite control an important role is played by the Casimir invariants $\Gamma_j$ ($j = 2,3,...,N$ for a $N$ level system).
The Casimir invariants of a state $\rho$ are related to the trace invariants $\rm{Tr}(\rho^j)$ ($j = 2, 3, ..., N$) and they cannot be altered by unitary transformation alone \ct{khaneja03, sauer13}.
For example, a two level system has a single Casimir invariant --its purity $P = \rm{Tr}\left(\rho^2 \right)$,
which remains unchanged under any unitary transformation. Consequently, any optimal strategy with the controls restricted to
unitary transformations only, would be to reach a state $\rho$ characterired by all Casimir invariants same as those of $\rho_{fp}$
in the minimum possible time. Following this we can apply a unitary pulse to reach the fixed point instantaneously. 
Clearly, any constrained control will at most
be as efficient as the results we present hereafter, based on the analysis we have presented previously for the case of
M dynamics~\ct{tannor99, mukherjee13}.

In what follows, we will analyse  Class A and B  channels independently.

{\bf Class A}: As shown in Fig.~\ref{schematic},
in the NM regime $d(t)$ goes to zero at $t = T_{F}$ when $\rho = \rho_{fp}$ and ${\cal L}(\rho_{fp}) = 0$. At the same time
we expect $|\gamma(t)| \to \infty$ at $t \approx T_{F}$ in order to have finite $\dot{\rho}(t) = \gamma(t) {\cal L}(\rho(t))$ even for
${\cal L}(\rho(t)) \approx {\cal L}(\rho_{fp}) = 0$, as is required for a non-monotonic $d$ of the form shown in Fig.~\ref{schematic}.
Notice that $\gamma(t)$ and hence the time $t = T_{F}$ at which $\gamma(t) \to \infty$ are in general independent of $\rho$.
Consequently any optimal control protocol which involves unitary transformation of $\rho(t)$ generated by $H_s(t)$ at earlier times
$t < T_{F}$ followed by non-unitary relaxation to $\rho_{fp}$ is expected to be ineffective in this case and we have
$T_{QSL} = T_{F}$. That is,
the gain (or efficiency) of optimal trajectory in the NM class is $R_{NM}^A=T_F/T_{QSL}=1$. One can easily see $T_F/T_{QSL} = 1$ implies absence of any speed up, whereas any advantage one gains by optimal control
can be quantified by $T_F/T_{QSL} > 1$.
On the other hand in the M limit  $\gamma(t)=\gamma_0$ is finite and constant, and
the system relaxes asymptotically to the fixed point in the absence of any control.
In this case we introduce an error tolerance $\epsilon \ll 1$, such that we say the target state is reached if
$|d(T_{F})| \leq \epsilon$. Clearly, $T_{F}$ increases with decreasing $\epsilon$ diverging to $T_{F} \to \infty$
in the limit $\epsilon \to 0$, as can be expected for finite $\gamma_0$.
Therefore the above argument of $|\gamma(t)| \to \infty$ at $t \approx T_{F}$
does not apply in this case and in general one can  expect the time of evolution to depend
on the initial state. Consequently the quantum speed up ratio $R_{M}^A$ can exceed $R_{NM}^A \approx 1$,
as is explicitly derived below in the case of a two level system in presence of
an amplitude damping channel.
Similar arguments apply also in the case when an additional unitary transformation is needed at the end of the evolution to reach $\rho_{fp}$, where $R_{M}^{A} \to \infty$ for $\epsilon \to 0$ \ct{mukherjee13}.

Our above result $R_{M}^A \geq R_{NM}^A$ can be expected to be valid in a more generic scenario with $\dot{\rho}(t) = \sum_k \gamma_k(t) {\cal L}_k (\rho(t))$ as well, where not all $\gamma_k$'s ($\neq 0$) are same, 
${\cal L}_k$'s are the time independent Lindblad generators and  
the unique fixed point $\rho_{fp}$ is defined by  ${\cal L}_k(\rho_fp) = 0$ for all $k$. In this case at least one of the $\gamma_k$'s can be expected to diverge at time $t = T_F$ in order to ensure Class A NM dynamics as shown in Fig (\ref{schematic}), 
thus making any optimal control ineffective as detailed above. We note that one can have dynamics with time dependent Lindblad generators and uncontrollable drift Hamiltonians acting on the system during the course of 
the evolution, in addition to the instantaneous control pulses, as well. The drift Hamiltonians can be expected to modify the Lindblad generators thus making the problem more complex;
however the analysis in this case is beyond the scope of our present work.

{\bf Class B:} Here we focus on systems of Class B where as already mentioned $\gamma(t)$ changes sign for $t_s < T_{F}$ with
$s = 1, \dots, N_s$. Clearly, in this case $|\gamma(t)|$ does not necessarily diverge for any $t$. Consequently the arguments presented above for class A fails to hold any longer and
the time of relaxation to the fixed point can in general be
expected to depend on $\rho(t)$ (and hence on $H_s$). Furthermore, it might be possible to exploit the NM effects such that even though $\dot{d}(t) > 0$ for $t_1 \leq t \leq t_2$ one can, by application of optimal control,
make sure that $\dot{\Gamma}_j(t) > 0$ and maximum $\forall~t,j$ (where we have assumed $\Gamma_j(t = 0) < \Gamma_{jf}$ $\forall~j$ and $\Gamma_{jf}$ denotes the $j$th Casimir invariants for the fixed point $\rho_{fp}$). This presents the possibility of
exploiting NM effects to achieve better control as opposed to the M dynamics, as is presented below for the case of a two level system in the presence of an amplitude damping channel.
 However we stress that
this is not a general result and explicit examples can be constructed where this is actually not true.

\subsection{Generalized amplitude damping channel \label{AD1}}

Let us now analyze in detail the generic formalism outlined above for the specific case of a two level system in contact
with NM amplitude damping channels of the two classes introduced before.

We consider the non-unitary dissipative dynamics described by the time local master equation 
acting on a  $2 \times 2$ reduced density matrix $\rho(t)$ of a qubit and we consider the time independent Linbladian ${\cal L}$ given by
\ba
{\cal L}(\rho(t)) &=& {\cal L}_1(\rho(t)) + e^{\beta}{\cal L}_2(\rho(t)), \non\\
{\cal L}_1(\rho(t)) &=& \left(\sigma_+\rho(t)\sigma_- - \frac{1}{2}\{\sigma_-\sigma_+, \rho(t)\}\right), \non\\
{\cal L}_2(\rho(t)) &=& \left(\sigma_-\rho(t)\sigma_+ - \frac{1}{2}\{\sigma_+\sigma_-, \rho(t)\}\right)\;,
\label{ADsuper}
\ea
with $\sigma_{\pm}$ being the raising/lowering qubit operators and $1/\beta$ gives the temperature of the bath.
The system evolution is given by Eq.~\ref{mastergen}, and we will focus on two different functional dependence of the parameter
$\gamma(t)$ corresponding to the Class A and B dynamics.
We will analyze the system evolution following the Bloch vector $\vec r$ representing the state $\rho = \left(I + \vec{r}.\vec{\sigma} \right)/2$ inside the Bloch sphere, where
the unitary part of the dynamics generated by $H_s$ induce rotations, thus preserving the
purity $P = (1 + |\vec r|^2)/2$. In contrast, in general the action of the noise  is expected to modify the purity as well.

{\bf Class A}:  An example of this class of dynamics is obtained under the assumption
\ba
\gamma(t) = \frac{2\lambda \gamma_0 \sinh \frac{tg}{2}}{g \cosh\frac{tg}{2} + \lambda \sinh\frac{tg}{2}}; ~~g = \sqrt{\lambda^2 - 2\gamma_0 \lambda}.
\label{JC}
\ea
In the above expression $\lambda$ and $\gamma_0$ are two positive constants whose ratio determines the bath behavior:
$\lambda > 2\gamma_0$ corresponds to a M bath, whereas $g$ becomes imaginary in the
limit $\gamma_0 > \lambda/2$ resulting in NM dynamics. In the NM limit of $\gamma_0 \gg \lambda, 1$
the bath time scale is determined by the product $\lambda \gamma_0$ and is independent of the specific form of the super-operator ${\cal L}$.
It can be easily seen that $\gamma(t)$ increases monotonically from $0$ to $\gamma(t) \to \infty$ at
\ba
\lim_{\gamma_0/\lambda \to \infty} T_{F} \approx \frac{\pi}{\sqrt{2\lambda\gamma_0}},
\label{Tfreenm}
\ea
where  $T_{F}$ is independent of the initial state and the system reaches the fixed point when $\gamma(t)$ diverges.
With this choice of $\gamma(t)$ in Eq. (\ref{JC}) the time scale is given by $\sqrt{2\lambda \gamma_0}$ in the NM
limit $\gamma_0 \gg \lambda$, while $\gamma(t) \approx \gamma_0$ sets the time scale in the M limit $\lambda \gg \gamma_0$.
Therefore the time taken to reach the fixed point can be expected to decrease as $1/\sqrt{\lambda\gamma_0}$ in the NM limit while it scales as $\sim 1/\gamma_0$ in the M limit.

As mentioned above, this form of $\gamma(t)$ arises in the damped Jaynes-Cummings model at absolute zero temperature, where one considers only a single excitation in the qubit-cavity system and Eq. (\ref{ADsuper}) reduces to 
${\cal L}(\rho(t)) = {\cal L}_2(\rho(t))$. However, here we consider a phenomenological form of the Lindblad generator (\ref{ADsuper}) with arbitrary $\beta$ to show the generality of our results.
In this context the parameter $\lambda$
in $\gamma(t)$ denotes the spectral width of the coupling to the reservoir, while $\gamma_0$ characterizes the strength of the coupling. 

In the absence of any control the qubit  relaxes to a fixed point
$\rho_{fp}$ characterized by the Bloch vector $\left(0,0,\frac{1 - e^\beta}{1 + e^\beta}\right)$ and the optimal control we analyze here aims
to accelerate the relaxation towards this state with unconstrained unitary control.  Following the strategy proposed above,
we look for the extremal speed  $v = \partial P/\partial t$ of purity change for every $r$.
For this model, the speed of change of purity is given by  $v(r,\theta,t) = -\gamma(t) (e^{\beta} - 1) r \left[\cos\theta + \frac{r}{2r_{fp}}\left(1 + \cos^2 \theta \right) \right]$ :
note that a positive (negative) $v$ denotes increasing (decreasing) purity. The two strategies differ slightly in case of
cooling or heating (i.e. the final purity is lower or higher than the initial one); but both cases correspond to applying unitary rotations at the beginning and at the end (for heating)
of the dynamical evolution, thus yielding a trajectory of the form Eq.~(\ref{JNEWEQ})  which is fully compatible with the CP requirement and which doesn't pose any problem in terms of
physical implementation (see discussion in Sec.~\ref{Rgen}). Specifically  we need to apply unitary control so that the system evolves along $\theta = \pi$ till the final purity is reached in the case of cooling, while
 $\theta = 0$ is the optimal path in the case of heating \ct{mukherjee13}, in agreement with a recent work on quantum speed limit in open quantum systems~\ct{deffner13}
(see Appendix~\ref{A1} for details).
\begin{figure}[t]
\begin{center}
\includegraphics[width= 9.4 cm,height = 4.5 cm, angle = 0]{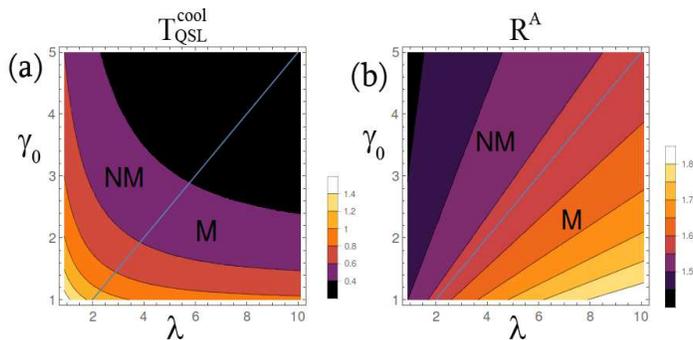}
\end{center}
\caption {(Color online) (a) Parametric plot showing variation of time $T_{QSL}^{cool}$ of reaching the fixed point with $\lambda$ and $\gamma_0$ for $\beta = 2$, $r_i = 0.5$ and $\epsilon = 0.01$. The Markovian (M)
and non-Markovian (NM) regions are separated by the blue line on the $\lambda - \gamma_0$ plane. (b)  Plot showing variation of quantum speed up ratio $R^A$ with $\lambda$ and $\gamma_0$ for
$\beta = 2$, $r_{xi} = 0.3$, $r_{yi} = 0$, $r_{zi} = 0.4$ and $\epsilon = 0.01$. $R^{A}$ saturates to $R^A_M \approx 2$ ($R^A_{NM} \approx 1$) in the extreme M (NM) limit.}
\label{Tqsl}
\end{figure}
Our analysis clearly shows that $T_{QSL}^{cool}$ decreases with increasing $\lambda$ as $\sim 1/\sqrt{\lambda}$ for small $\lambda$, finally saturating to $\lambda$ independent
constant values in the M limit ($\lambda \gg 1$). However, it would be misleading to conclude about the role of Markovianity on $T_{QSL}$ from this alone, since
both $T_{QSL}^{cool}$ and $T_{QSL}^{heat}$  decrease with increasing $\gamma_0$  as well. In particular, they scale as $ \sim 1/\gamma_0$ in the
M limit of small $\gamma_0$, while the scaling changes to $1/\sqrt{\gamma_0}$ for $\lambda/\gamma_0 \to 0$.
Indeed, the behavior depends more on the specific path in the $(\gamma_0, \lambda)$
plane  rather than whether the system is M or NM (see Fig.~\ref{Tqsl}a). However, if one analyzes the speedup
obtained by means of optimal control strategies, the scenario changes: in this case (as shown in Fig.~\ref{Tqsl}b) the ratio
$R^A = T_F/T_{QSL}^{cool}$ clearly distinguishes between M and NM dynamics with typical limiting values given by $\lim_{\epsilon \to 0, \lambda/\gamma_0 \to \infty} R_{M}^{A} \to 2$ and $\lim_{\epsilon \to 0,  \lambda/\gamma_0 \to 0} R_{NM}^{A} \to 1$ (see Appendix for details).
Finally, one can show that a control pulse can be considered to be instantaneous as long as its width in time $\delta t \ll 1/\left[\gamma_0 (e^{\beta} + 1) \right]$ in the M limit of $\lambda \gg \gamma_0$,
while in the NM limit $\lambda \ll \gamma_0$ one 
has $\delta t \ll 1/\sqrt{\lambda \gamma_0 (e^{\beta} + 1)}$.

{\bf Class B:} Finally, we investigate a particular case belonging to the class B dynamics and compare it to the previous case.
As in the previous case we consider a time evolution described by a master equation Eq.~(\ref{mastergen}) with ${\cal L}$ given by Eq. (\ref{ADsuper}); however for our present purpose  we formulate a $\gamma(t)$ given by
\ba
\gamma(t) = e^{-\zeta t} \cos(\Omega t)\;,
\label{nmgenII}
\ea
with $\zeta, \Omega$ being two positive constants satisfying the CP condition Eq.~(\ref{JNEWCON}). With this choice, in the absence of the control Hamiltonian,
NM effects manifest themselves for $(2n + 1)\pi/2 < \Omega t < (2n + 3)\pi/2$ for
integer $n \geq 0$ as $\gamma(t)$ changes
sign at $\Omega t = (2n + 1)\pi/2$, simultaneously altering the sign of $\dot{d}(t)$ to $\dot{d}(t) > 0$.
With a proper choice of parameters one can make $\gamma(t)$ (and hence $d(t)$) exhibit oscillatory dynamics for
$1 > d > 0$.
As for the previous example, also in this case the extremals of $v$ are independent of $\gamma(t)$ and determined
by $\mathcal{L}(\rho(t))$ only. Therefore they occur at exactly the same points  as for class A (\ref{JC}), i.e., at $\theta = 0, \pi$ and $\arccos (r/r_{f})$. In this case an
instantaneous pulse would correspond to its time width 
$\delta t \ll \rm{min}\{1/\zeta, 1/\Omega, 1/\left(e^{\beta} + 1 \right) \}$.
\begin{figure}[t]
\begin{center}
\includegraphics[width=8.0 cm,height = 3.4cm, angle = 0]{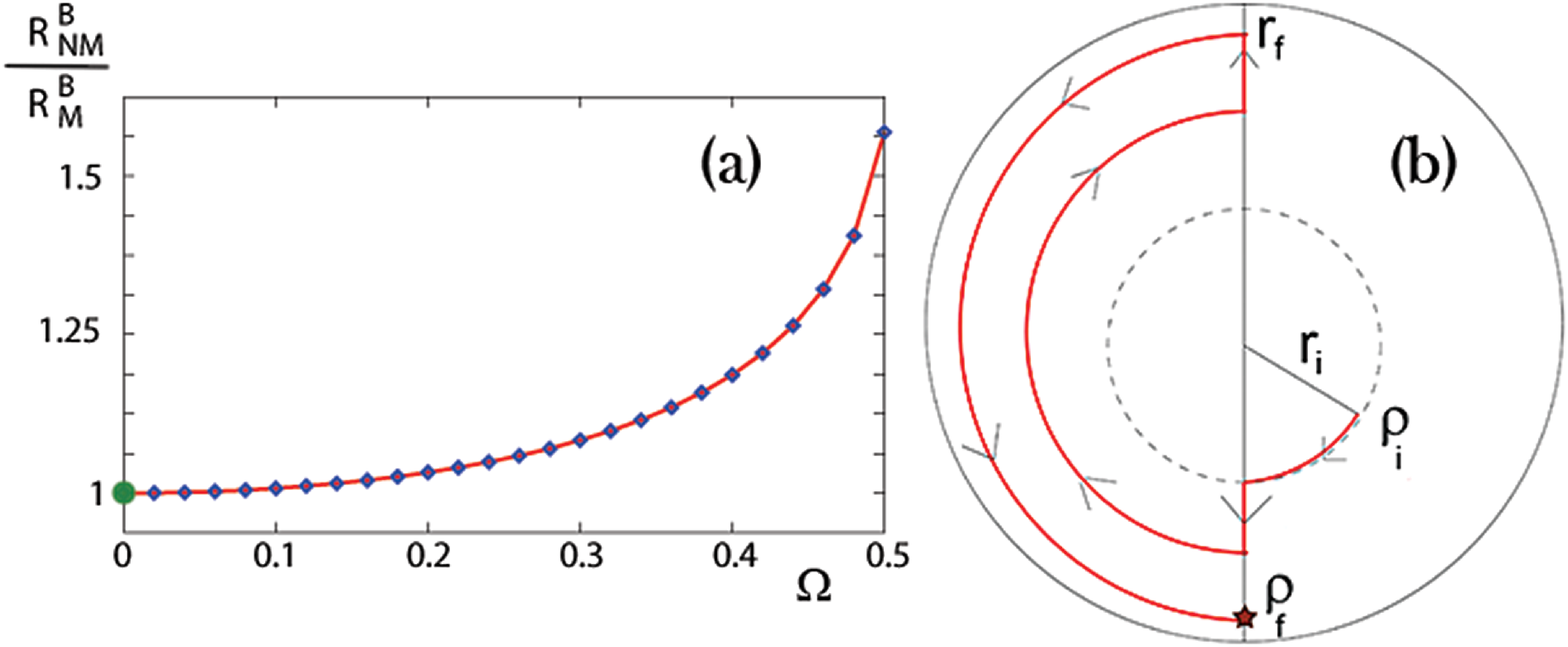}
\end{center}
\caption { (Color online) (a) Plot showing the variation of the ratio of the
gains $R_{NM}^B/R_M^B$ of the optimal trajectory as a function of $\Omega$ in case of Class B (cooling) with
$\gamma(t) = exp(- t) \cos(\Omega t), \beta = 2, \epsilon = 0.01$ and initial state given by $ r_{xi} = 0.3, r_{yi} = 0, r_{zi} = 0.4$. 
Clearly gain $R_M^B$ ($= 3.4$) in the M limit $\Omega = 0$, shown by the green dot, is less than that  ($R_{NM}^B$) in the NM limit
 $\Omega > 0$. (b) Schematic diagram showing optimal path in the $x-z$ plane of the Bloch sphere (red curve) in case of 
cooling a qubit in Class B (Eq. (\ref{nmgenII})), when we start from an arbitrary state $\rho_i$.
The fixed point $\rho_f$ is denoted by brown star.
}
\label{ADIIfig}
\end{figure}

The unconstrained optimal strategies are now modified as follows.
In the case of cooling, the optimal strategy
is to follow the path $\theta = \pi$ for $0 \leq t < T_{QSL}^{cool}$, during
which time the purity increases monotonically, where we have assumed the system reaches the target at
$t = T_{QSL}^{cool} < \pi/(2 \Omega)$ for simplicity. Consequently the 
optimal strategy demands a single pulse at time $t = 0$ only to make $\theta = \pi$, which corresponds to an evolution operator of the 
form ${\cal U}_{fin} \circ D \circ {\cal U}_{in}$. Clearly, this evolution is CPT as already discussed
in section \ref{intro} with $D$ depending on $\gamma_t$ and $\mathcal{L}$ (see Eqs. (\ref{ADsuper}) and (\ref{nmgenII})) and is thus possible
to implement physically.  In Fig.~\ref{ADIIfig}(a) we report the speedup obtained by such optimal strategy for different values of
$\Omega$ in Eq. (\ref{nmgenII}), where we have taken $\epsilon = 0.01$ large enough so that
$\Omega T_{QSL}^{cool} < \pi/2$. In this case we arrive at the
M limit by setting $\Omega=0$; as can be clearly seen the speed up in the NM limit is such that
$R_{NM}^{B} > R_M^{B}, \forall ~\Omega >0$ showing that there exist scenarios where NM effects can be exploited to improve the control
effectiveness. On the contrary, if the system does not reach the
target for $\Omega T_{QSL}^{cool} < \pi/2$, the optimal strategy changes: at $\Omega t = \pi/2$, $\gamma(t)$ and hence $v$ change sign, leading to 
decrease (increase) of purity for $\theta = \pi$ ($\theta = 0$).
Interestingly, we can take advantage of this effect by making $\theta = 0$ at $t = \pi/(2\Omega)$, where $v$ exhibits a maximum for $\gamma(t) < 0$. As mentioned before, application of a unitary pulse during the course of an evolution
may change the 
form of $\gamma(t)$ and $\mathcal{L}$ (see Eq. (\ref{fullevol})). 
However, for simplicity let us assume $v$ changes sign at $t = (2n + 1)\pi/2$ and assumes extremum values at $\theta = 0,\pi$ 
and $\arccos\left(-r/r_{fp}\right)$ even in presence of unitary control. One can easily extend our 
analysis to a more generic case where the simplifying 
assumptions do not hold by following the path of maximum (minimum) $v$ for cooling (heating). Let us first consider the case where the system reaches the target
$r_z = r_{f} - \epsilon$ at $t = T^{cool}_{QSL}< 3\pi/(2 \Omega)$. In such a scenario, as depicted in Fig.~\ref{ADIIfig} (b),
we let the system evolve freely for  $\pi/(2\Omega) < t \leq T_{QSL}^{cool}$, following which we take the system to $\theta = \pi$ and $r_z = -r_{f} + \epsilon$, thus obtaining the desired goal.
Clearly, an optimal path exists in case of the class B non-Markovian channel,
which if possible to be followed by application of suitable
unitary controls, helps in cooling and in particular might
make it possible to reach the fixed point in finite time (if $\epsilon = 0$). Generalization to the case
$\Omega T_{QSL}^{cool} > 3\pi/2$ where multiple $\pi$ rotations are needed is straightforward.  However, we emphasize that the strategy presented above for $\Omega T_{QSL}^{cool} > \pi/2$ follows an evolution operator of the 
form Eq.~(\ref{fullevol})
with unitary pulses applied at intermediate times.
Consequently our analysis  gives a lower bound 
to $T_{QSL}^{cool}$ only, achieved by following the optimal path
shown in Fig. \ref{ADIIfig}(b), for which at present we do not have
any implementation strategy.

Finally, we address the  problem of heating the system in the shortest possible time, which amounts to minimizing $v$ $\forall~t$.
Therefore, the optimal path dictates to set $\theta = 0$ at $t = 0$ and then let it evolve freely till $\Omega t = \pi/2$,
where $\gamma(t)$ and hence $v$ change sign.
Unfortunately in contrast to the cooling problem, now $v > 0$ $\forall \theta$
making it impossible to take advantage  of the NM effects to accelerate the evolution.
However, even in this case one can always minimize the unwanted
effect of information backflow for $\gamma(t) < 0$ by increasing
$\theta$ to  $\theta = \arccos (- r_{fp}/r)$ where $v$ has a minimum.

\section{Conclusion \label{concl}}
We have studied the effectiveness of unconstrained optimal control of a generic quantum system in the presence of a non-Markovian dissipative bath.
Contrary to common expectations, the speedup does not crucially depend on the Markovian versus non-Markovian divide, but rather on the specific details of the non-Markovian evolution.
We showed that the speed up drastically depends on whether the system dynamics is monotonic
or not before reaching the fixed point for the first time, as determined by the trace distance to the fixed point (class A and class B dynamics respectively).
Indeed,  in the former case, the speed up obtained via optimal control is always higher in presence of
a Markovian bath as compared to a non-Markovian one, while the reverse can be true in the latter case.
Finally, we have presented some specific examples of these findings for the case of a two level system subject to an amplitude damping channel.
Note that, in the more realistic scenario where one can apply control pulses of finite strength only, the presented results serve as theoretical
bounds to the optimal control effectiveness.\\

{\em Acknowledgements --} The authors acknowledge Andrea Mari, Andrea Smirne, Alberto Carlini and Eric Lutz for helpful discussions. We ackowledge support from the Deutsche Forschungsgemeinshaft (DFG) within the
SFB TR21 and
the EU through EU-TherMiQ (Grant Agreement 618074), QUIBEC, SIQS, the STREP project PAPETS and QUCHIP.

\section{appendix}
\label{A1}

\subsection{Class A optimal strategy}

Analysis of $v(r,\theta,t) = -\gamma(t) (e^{\beta} - 1) r \left[\cos\theta + \frac{r}{2r_{fp}}\left(1 + \cos^2 \theta \right) \right]$ shows
the optimal strategy in case of cooling
is to apply a unitary pulse at $t = 0$ so as to rotate $\vec{r}$ to $\theta = \pi$.
Following this we switch off the control and allow the qubit to relax by the application of the dissipative bath alone for a time $t = T_{QSL}^{cool}$, till it reaches $r(T_{QSL}^{cool}) = r_{fp} - \epsilon$.
In contrast, while considering the problem of minimizing the time taken to heat the qubit, the optimal strategy is to first rotate the Bloch vector to $\theta = 0$.
As before, we then turn off the unitary control and let it relax till it reaches $(0, 0, r_{fp})$, following which we apply a second unitary pulse to take the system to the fixed point $\vec{r}_{fp}$ \ct{mukherjee13}.

We use the optimal strategy formalism presented above  to arrive at the minimum time $T_{QSL}^{cool}$ needed
to cool the system in the different limits.
The time for cooling in the M limit $\lambda/\gamma_0 \to \infty$ is given by
\ba
\lim_{\lambda/\gamma_0 \to \infty}T_{QSL}^{cool} \approx \frac{1}{\gamma_0 (1 + e^{\beta})} \ln \frac{r_{fp} - r_i}{\epsilon},
\label{tcoolm}
\ea
whereas the same in the NM limit is
\ba
\lim_{\lambda/\gamma_0 \to 0}T_{QSL}^{cool} \approx \sqrt{\frac{2}{\lambda \gamma_0}}\left[\frac{\pi}{2} -  \left(\frac{\epsilon}{r_{fp} - r_i} \right)^{\frac{1}{\left[2(\exp(\beta) + 1)\right]}}\right].
\label{Tnmc}
\ea
On the other hand, following the optimal strategy to heat the qubit one gets
\ba
\lim_{\lambda/\gamma_0 \to \infty}T_{QSL}^{heat} \approx \frac{1}{\gamma_0 (1 + e^{\beta})} \ln \left[\frac{r_{fp} + r_i}{2 r_{fp} + \epsilon}\right]
\label{tmh}
\ea
in the M limit, while in the NM limit it is
\ba
\lim_{\lambda/\gamma_0 \to 0} T_{QSL}^{heat} \approx \sqrt{\frac{2}{\lambda \gamma_0}} \cos^{-1}\left[ \left(\frac{2 r_{fp} + \epsilon}{r_i + r_{fp}}\right)^{\frac{1}{2(exp(\beta) + 1)}}\right].
\label{Tnmh}
\ea
As can be easily seen,
one needs to consider a non-zero $\epsilon$ in order to keep the time of cooling finite in the M limit while we can set it exactly to $0$ in the other cases and yet reach the
fixed point in finite time.

In contrast to the results derived above, the advantage one gains by application of optimal control presents a completely
 different picture. As before, one can understand this from the the quantum speed-up ratio $R^A = T_{F}/T_{QSL}$.
In the M limit $\lambda/\gamma_0 \to \infty$, $\gamma(t) \approx \gamma_0$ one gets
\ba
T_{F} \approx \frac{2}{\gamma_0 \left(1 + e^{\beta}\right)} \ln \frac{|r_{xi}|}{\epsilon},
\label{Tfree}
\ea
where we have assumed $r_{xi} \gg \epsilon$, which is typically the case for $\epsilon \to 0$.
In the case of cooling, as can be seen from Eq. (\ref{tcoolm}) and Eq. (\ref{Tfree}), $R_{M}^{A} \approx 2$ in
the M limit as long as $r_{xi}, (r_{fp} - r_i) \gg \epsilon$ and $\epsilon \to 0$. On the other hand, the NM limit of $\lambda/\gamma_0 \to 0$ yields
\ba
T_{F} \approx \sqrt{\frac{2}{\lambda \gamma_0}} \left[\frac{\pi}{2} - \left(\frac{\epsilon}{r_{x0}} \right)^{2/(\exp(\beta) + 1)} \right],
\label{freenm}
\ea
thus reducing the gain to $R_{NM}^{A} \approx 1$ in the limit of $\epsilon \to 0$ (see Fig. (\ref{Tqsl}b)).
In case of heating we have
\ba
\lim_{\epsilon \to 0} R_M^{A} \approx 2\frac{|\ln \epsilon|}{\ln \left[\frac{r_{fp} + r_i}{2 r_{fp}} \right]}
\label{gainm}
\ea
in the M limit, while the same in the NM limit is
\ba
\lim_{\epsilon \to 0} R_{NM}^{A} \approx \frac{\pi/2}{\cos^{-1} \left[\left(\frac{2 r_{fp}}{r_i + r_{fp}} \right)^{\frac{1}{2(\exp(\beta) + 1)}} \right]},
\ea
which again implies the gain is much higher in the M limit for $\epsilon \to 0$.

\end{document}